\documentclass[10pt]{article}
\usepackage{booktabs}
\usepackage{siunitx}
\usepackage{float}
\usepackage{subcaption}
\usepackage[margin=1.0in]{geometry}
\usepackage{amsmath}
\usepackage{xcolor}
\usepackage[colorlinks,allcolors=black]{hyperref}
\usepackage[numbers]{natbib}

\usepackage{graphicx}
\usepackage{comment}

\title{Valuing Post-Revenue Biopharmaceutical Assets with Pfizer's Current Portfolio as a Case Study}

\author{Yongzhuo Chen  \\
	Yale University  \\
	\and 
	Yixuan Liang \\
	Yale University \\
	 \and
    Yiran Liu \\
    Yale University \\
    \and
    Brian Hobbs \\
    The University of Texas at Austin \\
    \and
    Michael Kane \\
    Yale University \\
  }

\begin{document}

\maketitle

\begin{abstract}
This research paper addresses the critical challenge of accurately valuing post-revenue drug assets in the biotechnology and pharmaceutical sectors, a key factor influencing a wide range of strategic operations and investment decisions. Recognizing the importance of reliable valuations for stakeholders such as pharmaceutical companies, venture capitalists, and private equity firms, this study introduces a novel model for forecasting future sales of post-revenue biopharmaceutical assets. The proposed model leverages historical sales data, a resource known for its high quality and availability in company financial records, to produce distributional estimates of cumulative sales for individual assets. These estimates are instrumental in calculating the Net Present Value of each asset, thereby facilitating more informed and strategic investment decisions. A practical application of this model is demonstrated through its implementation in analyzing Pfizer's portfolio of post-revenue assets. The model's effectiveness is evidenced by its ability to project Pfizer's market capitalization—defined as the product of its share price and the number of outstanding shares whose accuracy achieves a margin of error within 10\%. This precision highlights the model's potential as a valuable tool in the financial assessment and decision-making processes within the biotech and pharmaceutical industries, offering a methodical approach to identifying  investment opportunities and optimizing capital allocation.
\end{abstract}

\section{Keywords}\label{background}

Biotechnology post-revenue asset valuation, pharmaceutical portfolio valuation

\section{Introduction}

Accurate valuation of post-revenue drug assets plays an instrumental role in a broad range of operations and decisions in the biotech and pharmaceutical sectors. Investment decisions heavily depend on accurate valuations. They inform decisions made not only by pharmaceutical companies but also by a range of stakeholders including venture capitalists, private equity firms, and others. An accurate valuation provides a reliable forecast of the potential return on investment, which directly influences where and how capital is allocated within the industry. It forms the basis of strategic investment decisions, making it possible to identify the most promising opportunities and understand which investments are more likely to yield a satisfactory return.

Valuation plays a key role in the due diligence process of Mergers and Acquisitions. During this phase, buyers conduct a comprehensive analysis of both the financial and operational aspects of the target company. The valuation serves as a crucial reference point against which the findings of the due diligence are evaluated. Any significant disparities identified may trigger further investigation or necessitate adjustments to the deal terms. Overestimating or underestimating the value of a pharmaceutical asset will result in the buyer paying more than necessary or the seller receiving less than their potential earnings. Therefore, an accurate valuation is imperative to ensure fairness and equity in transactions for all parties involved. In addition, given the inherent uncertainties in drug development and approval processes in risk management, a proper valuation provides a realistic appraisal of potential risks and rewards, aiding in strategic planning and contingency creation. Moreover, it facilitates effective prioritization within a company's drug portfolio, empowering firms to make informed decisions about resource allocation based on the projected value of individual assets. Accurate valuation also becomes paramount in regulatory and legal scenarios involving patent disputes, litigation, or other regulatory matters, where a drug asset's value might be a critical piece of evidence. Lastly, precise valuation reinforces market confidence and upholds the credibility and integrity of the biopharma sector. It serves as a bulwark against asset price inflation, safeguarding financial stability in the market by ensuring that the disclosed value aligns with the real worth of the asset.

After a drug has gained regulatory approval and becomes a monetizable asset, there are clinical risks that could influence its valuation. Post-market safety issues may emerge that lead to product recalls or the imposition of black-box warnings. Securing reimbursement from insurers or government healthcare programs can present challenges, even for an approved drug. The level of reimbursement can greatly influence the uptake of the drug and hence its revenue generation. Rapid advancements in technology and changes in medical practice can impact the relevance and efficacy of a drug, which can dramatically affect its valuation. A drug that was once deemed revolutionary can quickly become obsolete due to the advent of a new technologies or shifts in medical practice. Finally, when a patent expires, it can lead to the introduction of generic competition, which in turn can lead to a significant reduction in revenues of the branded drug. While the timing of patent expiration is generally known, predicting the impact of generic competition on sales is itself a complex task and can be dependent on factors including the complexity of the manufacturing process, the competitive landscape of the generic market, market exclusivity periods, potential pricing strategies by generic competitors, the breadth and loyalty of the customer base for the branded drug, the marketing and distribution capabilities of the generic drug manufacturers, and potential regulatory changes. Understanding these dynamics requires deep industry knowledge and careful analysis of market trends, making it a challenging aspect of post-revenue drug asset valuation.

Valuing post-revenue drug assets is a complex undertaking due to several interrelated factors. The pharmaceutical and healthcare sectors are characterized by a high degree of volatility. Decisions at the political level, economic cycles, technological breakthroughs, and public health crises can all trigger significant fluctuations in the market. Such unpredictability further complicates the task of accurately forecasting the performance of a drug. An accurate valuation requires a clear understanding of the market size, penetration rates, competition, pricing, and adoption rates by doctors and patients. Adding to the complexity is the dynamic nature of the regulatory environment. Changes in regulations around pricing, alterations in patent laws, or shifts in approval processes can all unexpectedly impact a drug's future revenue.

Valuing pharmaceutical assets and the portfolios they comprise has seen some modest recent activity. Kellogg et al, \cite{kellogg2000real} proposed a real-options approach to valuing Small Life Sciences Companies (SLSC) which Kane et al.\cite{kane2021} distinguished from Biotechnology companies with the latter tending to hold both pre- and post- revenue assets and the former holding pre-revenue assets almost exclusively. Stasior \cite{Stasior2018} 
recently propose a risk-adjusted, net present value (rNPV) model for individual assets that incorporates the unique risks and rewards associated with post-revenue pharmaceutical assets. Michaeli et al. \cite{michaeli2022valuation} evaluated the association of Biopharma company valuation with the lead drug’s development stage, orphan status, number of indications, and disease area and estimate annual returns entrepreneurs and investors can expect from founding and investing in drug development ventures.

For post-revenue valuations specifically, Ledley et al. \cite{ledley2020profitability} provide a comparison of pharmaceutical companies with other types of companies, finding that they are more profitable than other companies. The literature becomes sparse beyond this point, with almost no studies examining either the valuation of individual drug assets and uncertainty in those estimates or the valuation of portfolios as it relates to the market price of Biotech or Pharmaceutical companies. This paper fills in this gap by proposing and validating a model for future revenue for individual assets based on publicly available sales taking into account the process of market penetration, the inevitable incidents like loss of intellectual property rights, and competition for an indication, with the assumption that the incidents will cause the symmetrical decrease of the sales. An Epidemiology Sales Prediction Model also aims to predict the sales of health-related products by integrating epidemiological data such as incidence and prevalence rates of certain diseases, demographic data, healthcare policies, and other relevant factors that might impact the sales of these products\cite{Soyiri2012}. One drawback of these models is their heavy dependency on accurate epidemiological data\cite{Bakhta2020-bx}; however, obtaining high-quality data can sometimes be a challenge. Another drawback is the inherent complexity of the model, which can make it difficult to understand and use, especially for individuals not familiar with epidemiological principles. This complexity also may compromise the robustness of the model. The general and straightforward prior assumption in our model can substitute for the complex scenarios in epidemiology well. 

This paper presents a model designed to estimate future sales of post-revenue assets using their historical sales data. This data, known for its high quality, is readily accessible in the company's financial records. The model generates distributional estimates of each asset's cumulative sales, aiding in the calculation of the asset's Net Present Value (NPV). Applied to Pfizer's portfolio of post-revenue assets, this model accurately projects the company's market capitalization—defined as the product of its share price and the number of outstanding shares—to within a 10\% margin.

The paper proceeds as follows. Section \ref{sect:sales_curve} describes the sales curve in the classic Product Life-cycle \cite{ProductLifeCycle} as it applies to the pharmaceutical industry. From this description, a model is proposed in Section \ref{sect:model} taking into account the market demand for a therapy, limits on intellectual property rights, and competition. The model is validated using Pfizer's current drug assets sales in Section \ref{sect:validation}. Section \ref{sect:posthoc} provides to post-hoc analyses of the model with first indicating that the model accuracy is not enhanced by auxiliary information related to the therapy or indication space and the second identifying differences in uptake of therapies by their indication class. Section \ref{sect:portfolio} aggregates the individual asset valuations and their uncertainties to provide an estimate of Pfizer's current post-revenue portfolio value and compares that estimate with Pfizer's current market capitalization. The paper concludes with a discussion in Section  \ref{sect:discussion} which proposes extensions to revenue planning and to pre-revenue valuation.

\section{The pharmaceutical asset sales cycle} \label{sect:sales_curve}

The pharmaceutical sales curve for drug assets for fits into the classic product life cycle framework for analyzing the performance of a product and identifying appropriate actions to maximize its success \cite{ProductLifeCycle}. The life cycle framework describes the stages that a product goes through from its introduction to the market until its eventual decline and withdrawal. It consists of four distinct stages: introduction, growth, maturity, and decline. These stages provide a framework for analyzing the performance of a product and identifying appropriate actions to maximize its success. We note that this the product life cycle in pharmaceuticals has been examined before \cite{Product_lifecycle_management_pharmaceuticals}. However, here we focus on post-revenue and sales.

\begin{figure}[H]
    \includegraphics[width=\textwidth]{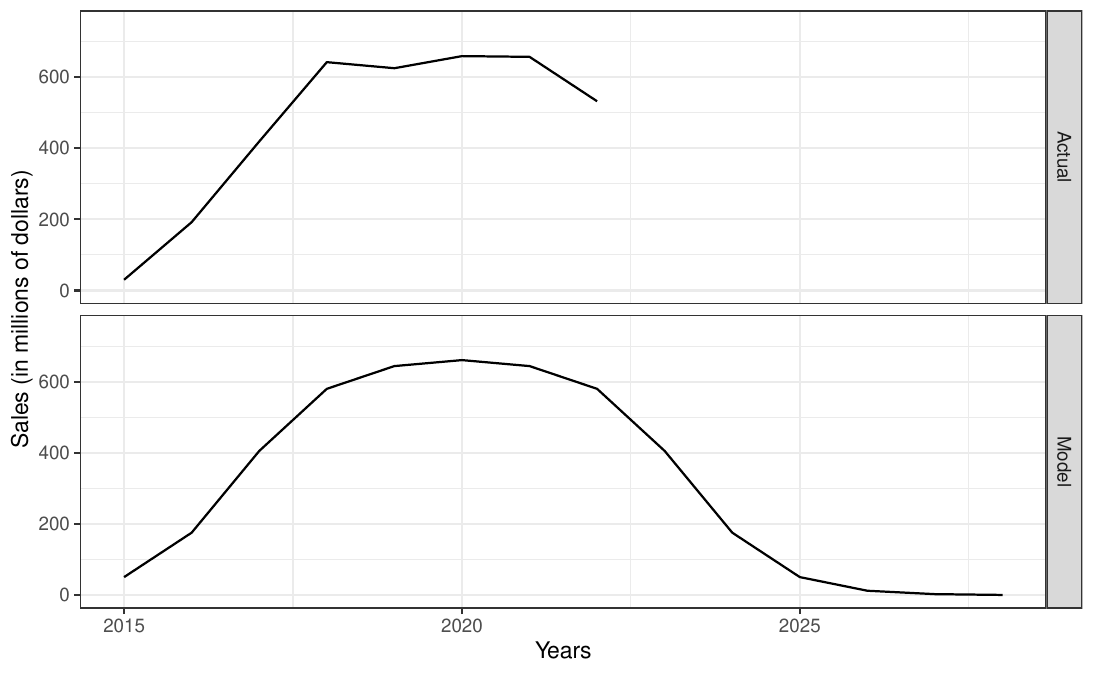}
    \caption{World-wide sales of Pfizer's Inflectra drug asset (above) and it's idealized (estimated) sales curve (below).}
    \label{vis:sales}
\end{figure} 

Consider the world-wide sales of Pfizer's Inflectra asset shown in the upper half of Figure \ref{vis:sales}. This is chimeric monoclonal antibody, sold under the brand name Remicade among others, targeting autoimmune indications includeing Crohn's disease, ulcerative colitis, rheumatoid arthritis, ankylosing spondylitis, psoriasis, psoriatic arthritis, and Behçet's disease. The drugs introduction to the market started after it's approval from authorities such as the U.S. Food and Drug Administration (FDA), the European Medicines Agency (EMA), and the Japanese Pharmaceuticals and Medical Devices Agency (PMDA). 

After approval, drug assets move into the growth stage as there is demand for the drug in the disease population and the use of the drug is normalized to caregiver by educational material provided by the drug company and, where it is legal, direct marketing to individuals. Inflectra was first monetized as multi-indication drug in 2015 and exhibited a precipitous uptake in the global market. This expedited entry into the growth phase can be attributed to a strategic confluence of comprehensive educational initiatives and targeted marketing campaigns. These measures effectively elucidated the drug's potential to both caregivers and patients, underpinning its swift adoption. Furthermore, Inflectra's broad therapeutic utility across numerous autoimmune indications facilitated its penetration into a large market segment. This ability to address diverse, previously unmet clinical needs served to amplify its commercial success.

Drug assets reach maturity and sales saturation when it has been accepted as the standard of care, it reaches a significant portion of the disease population, and the therapy is not expanded into new indications. For Inflectra, saturation occurred in 2018, where annual sales where approximately \$642 million world-wide and continued through 2021 where sales were approximately \$657 million.

The decline stage can be initiated for three reasons. The first can occur because of a loss of Intellectual Property (IP) protection. In most countries, the standard length of patent protection for a new drug is 20 years from the date of filing the patent application. However, the actual time a drug is protected and marketed exclusively by a company can be considerably less than this due to the time it takes for drug development and regulatory approval. However, many countries have provisions for extending the term of a patent beyond 20 years under certain conditions, often referred to as patent term extensions or supplementary protection certificates. These extensions are usually granted to compensate for the time spent in clinical trials and the regulatory approval process. After losing IP protection, many assets can be manufactured at much lower cost as a generic. The second reason for decline is that new treatments with improved efficacy or safety profiles may emerge, reducing the market demand for the existing drug. Advances in the biomedical field are constant, and it's not uncommon for newer therapies to exhibit enhanced efficacy or better safety profiles. These novel treatments may provide patients with similar therapeutic benefits but fewer side effects, or they may offer a greater degree of symptom relief or disease control. The third reason for a decline in drug asset sales is the availability of a biosimilar, a biological product that is highly similar to and has no clinically meaningful differences from an existing FDA-approved reference product. Biotech companies developing biosimilars stand to benefit from substantial revenue potential due to the large market for biologics and growing demand for cost-effective alternatives. The development of biosimilars, while complex, is generally less expensive than creating new biologics, providing a balance of lower upfront costs and favorable profit margins. Additionally, support from healthcare payers for more affordable treatment options and potential government incentives further enhances the financial appeal of biosimilars.

Inflecta is itself a biosimilar developed to rival Janssen Biotech Inc.'s Remicade (patented in 2001 \cite{le2001anti}) and initially capitalized on being one of the first biosimilars in this therapeutic category. However, the landscape has evolved with the U.S. approval of other biosimilars such as Samsung Bioepis's Renflexis in 2016, Amgen's Avsola in 2019, and Novartis's Zessly in 2018. These competitors, each offering similar therapeutic profiles, have significantly expanded the market supply. This growth in competition, through an enriched biosimilar portfolio, has led to a dilution of Inflectra's market share.

\section{Model} \label{sect:model}

In this section, we propose a model to predict future sales of an asset based on existing sales data. Product lifecycle models have been proposed before for a variety of scenarios our case fits most readily into the 3-parameter model \cite{ProductLifeCycle} written as:
\begin{equation} \label{threeparam}
y(t) = a + bt + ct^2
\end{equation}
where $y$ denotes sales, $t$ denotes time, and $a$, $b$, and $c$ are parameters that need to be calibrated to the sales curve. However, this model has several shortcomings in the context of biopharmaceutical assets that make it less appealing for prediction.

First, it does not directly integrate known parameters, including when an asset will lose IP protection. This value in particular signifies a latest time at which we expect to see the initiation of the decline phase. The loss of IP protection is often followed by a more rapid and abrupt drop in sales than what is predicted by the quadratic equation due to the surge of generic versions entering the market. The 3-parameter model's inability to effectively account for this key event limits its usefulness in making accurate sales projections for biopharmaceutical assets, especially in the later stages of their lifecycle. Consequently, any model aiming to better predict future sales of such assets needs to factor in these IP protection expiry dates and the consequent shifts in sales trends. Second, other models tend to model saturation as a point in the lifecycle. Real-world scenarios often demonstrate that drug assets can reach and maintain a saturation state for extended periods. In the case of pharmaceutical drugs, they may stay saturated due to factors such as a lack of alternative treatments or ongoing demand from a specific patient population. Third, the traditional models exhibit limitations in its exrapolative capabilities. This is especially true in the scenario where we would like to predict future sales from relatively data at the beginning of the lifecycle. Fourth, traditional models tend not to account for market uncertainty and distributional outcomes. In reality, various factors can introduce unpredictability into sales figures, such as changes in consumer preferences, competitive dynamics, or unforeseen events like economic downturns or regulatory shifts. The model's inability to provide distributional results hampers accurate decision-making and planning. Finally, the traditional 3-parameter models lack clear interpretability. They are merely mathematical coefficients without direct connections to underlying market dynamics or meaningful implications for decision-making. As a result, calibrating these parameters requires arbitrary estimation, diminishing the model's value for practical applications.

\begin{equation} \label{eqn:model}
y(t) = \frac{s \times 1_{\{t <= t_s\}}}{1 + e^{\beta_0 + t \beta_1}} + \frac{s \times 1_{\{t > t_s\}}}{1 + e^{\beta_0 + 2t_s - t \beta_1}}
\end{equation}

We proposed the pharmaceutical asset sales model shown in Equation \ref{eqn:model}, which accounts for market uncertainty, rapid transitions, saturation periods, and provides reliable distributional outcomes. In this case $y(t)$ is the sales at time $t$, $s$ is the saturated value of sales, $1_{\{x\}}$ denotes the indicator function over the set $x$, $t_s$ is the year corresponding to half-way through the sales cycle, $\beta_{0}$ and $\beta_{1}$ is the parameter the model going to fit, $t$ is the year. The model can be regarded as a piece-wise, symmetric logistic regression, with an extra parameter $s$ specifying the sales at saturation. The left side of the summand in Equation \ref{eqn:model} models the first half of the sales cycle and the left models the second half. The parameters $\beta_{0}$ and $\beta_{1}$ control the offset and transition to saturation.

It may be noted that the model is symmetric in $t_s$ meaning that the rate of decline will estimated as equal to the rate of growth. In practice this is not always the case, for examples, for small-molecule therapies, which have relatively simple manufacturing processes, the drop off may be immediate and precipitous. Other drugs, with complex manufacturing processes, may see slow decline if competitors are not able to easily integrate their production. However, in Section \ref{sect:validation} we show that this simplification is still able to achieve accurate sales predictions.

When predicting future sales, we assume that we have sales through some year either before or after saturation has been reached $\{y(t_0), y(t_1), ..., y(t_k)\}$ and we would like to predict sales through the remaining years over which the asset is monetized, $\{\hat{y}(t_0), \hat{y}(t_1), ..., \hat{y}(t_k)\}$. This implies estimates of $s$, $\beta_{0}$, $\beta_{1}$, and $t_s$ and $s$ if it has not already reached saturation. Our estimates of the distributions of these parameters will induce uncertainty that can then be characterized in the final estimates of cumulative future sales. In practice, for pre-saturation sales data, the majority of the uncertainty comes from estimates of $t_s$. 

Estimation of the parameter occurs in several two steps. First, estimation of the model scale parameters $s$ is determined by finding
\begin{equation} \label{eqn:smin}
    \hat{s} = \underset{s}{\operatorname{argmin}} \sum_{t_k = 0}^{t_{\text{max}}} \left(\frac{s}{1 + e^{\beta_0 + t_k \beta_1}} - y(t_k) \right)^2
\end{equation}
where $t_{max} = \{t : \text{max}(y(t))\}$. In practice this is performed fixing a candidate value $s$, and then the fraction in Equation \ref{eqn:smin} as well as the correspond $y(t_k)$ value. After the normalization, a logistic regression is performed to estimate $\beta_0$ and $\beta_1$. The process continues, refining the values of $s$ until convergence. If the sales are not already declining then, in the second step $t_s$ is selected by taking the time at which the asset loses IP protection minus a draw from an exponential distribution, truncated at the difference between time the asset losses IP and $t_max$. This step can be resampled and the variance in the parameter is used to model market uncertainty in the estimate.

\section{Model validation using Pfizer sales} \label{sect:validation}
To validate the model, we use revenue data taken from Pfizer's financial reports from 2006 through 2022. The  dataset comprises revenue data extracted from Pfizer's Financial Reports spanning from 2006 to 2019, specifically focusing on the revenues generated by several key pharmaceutical products. Additionally, information from the Annual Report on Form 10-K for the years 2020 to 2022 has been incorporated, primarily sourced from sections such as "Significant Product Revenues" and "Revenues—Selected Product Discussion." In addition, we have incorporated the patent expiration date for each asset as well as several other features including the number of phase 3 trials the asset was used as treatment for, the indication type targeted by each asset, and the total number of conditions targeted. Validation proceeds in the following two subsections as follows. First, we will hold out sales in later years and report the out-of-sample accuracy of the holdout data in terms of the MSE. Second, we will test the null hypothesis that the predictive information needed to estimate future sales is contained in the curve and that, as a result, information including number of phase 3 trials, indication type, and total number of conditions does not increase the model accuracy. Our results show that the model is accurate in terms of out-of-sample prediction accuracy and that the incorporation of the state features does not enhance prediction accuracy.

We note that in this section, we did not incorporate Cormirnaty and Paxlovid into the validation, Pfizer's two COVID-19 related assets. The sales of Cormirnaty exhibited a rapid increase during its initial year of availability, followed by a subsequent deceleration in its second year. Paxlovid also experienced a significant surge in sales, rising from 76 million to 18.9 billion units. Theses two remarkable increases can be attributed to the overwhelming demand for effective treatments and preventive measures against the COVID-19 pandemic. The urgent need for solutions during this global health crisis contributed significantly to the unprecedented growth of Paxlovid's sales. As the pandemic subsides, sales are expected to decline, potentially bypassing the saturation stage. This assumption is supported by the declaration made by the World Health Organization (WHO) chief, stating an end to COVID-19 as a global health emergency. 

For the remaining drugs in Pfizer's portfolio we rely on the proposed model to estimate sales identify the saturated year for these drugs. Here, we examine the expected cumulative returns for each asset. In this section, we will provide examples on two specific assets: Eliquis and Ibrance.

\begin{figure}[htbp]
  \centering
  \begin{subfigure}[b]{0.45\textwidth}
    \includegraphics[width=\textwidth]{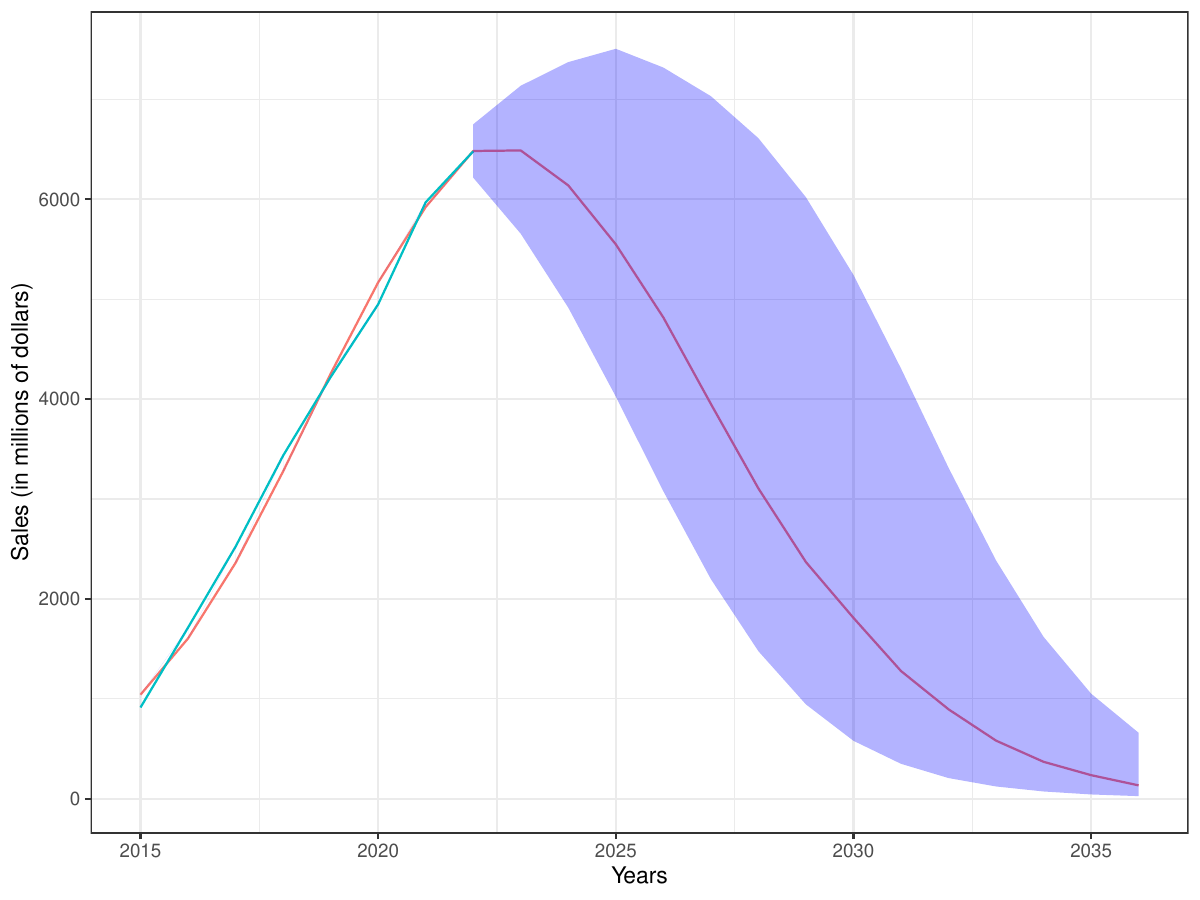}
    \caption{Eliquis}
    \label{fig:subplot_Eliquis_1}
  \end{subfigure}
  \hfill
  \begin{subfigure}[b]{0.45\textwidth}
    \includegraphics[width=\textwidth]{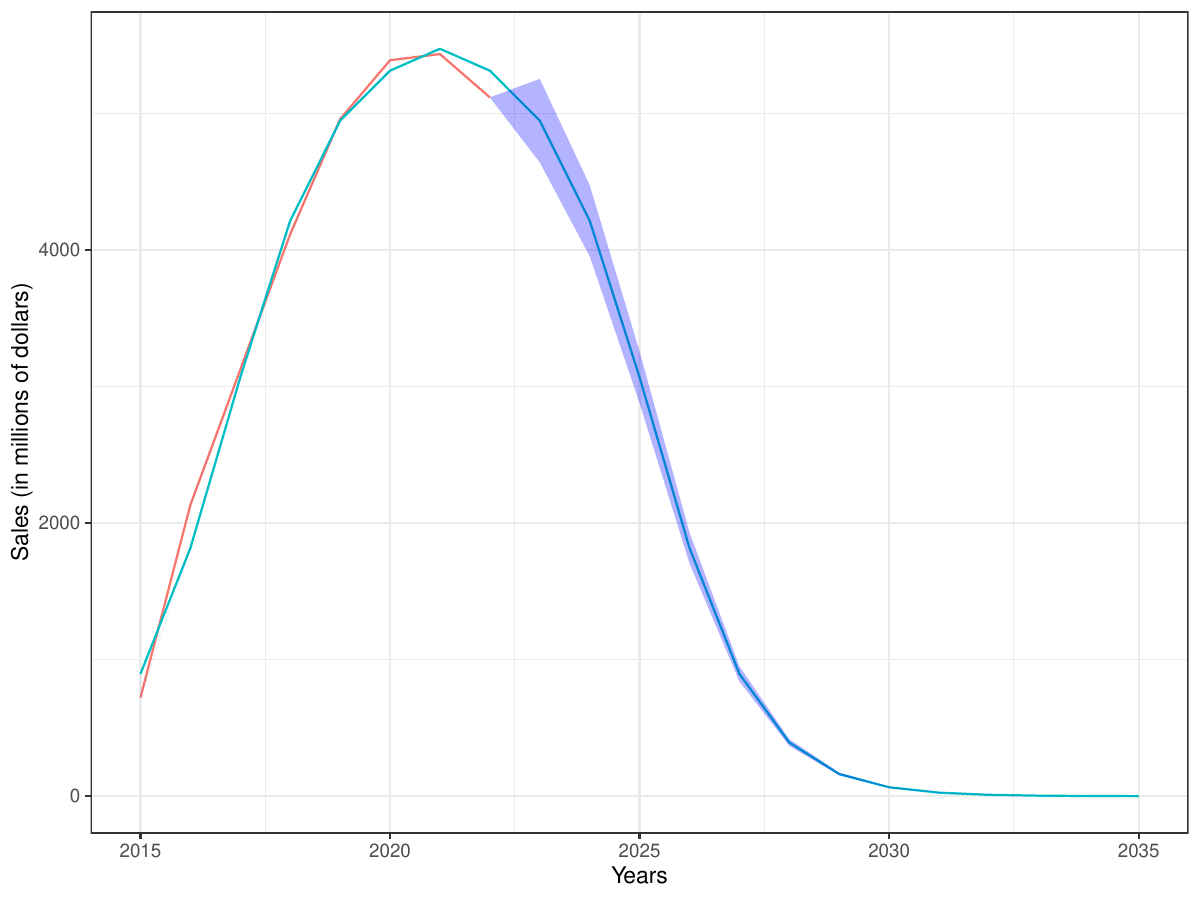}
    \caption{Ibrance}
    \label{fig:subplot_Ibrance_1}
  \end{subfigure}
  
  \caption{Figure of Eliquis and Ibrance}
  \label{fig:9subplots}
\end{figure}

For Eliquis, shown in Figure \ref{fig:subplot_Eliquis_1}, the model indicates that sales have yet reached saturation. Each of the scenarios where saturation is reach through the year the asset loses IP are shown as vertical facets. The cumulative returns in each of these scenarios are \$55.4 billion, \$68.6 billion, \$82.4 billion, and \$96.1 billion respectively. The corresponding probabilities for these returns are 0.3981, 0.332, 0.192, and 0.078, respectively. The expected total sales for Eliquis then estimated to be \$67.5 billion.

For Ibrance, shown in Figure \ref{fig:subplot_Ibrance_1}, the model suggests that its sales have already reached sales saturation the seventh year from its release(2020). Consequently, the model provides a specific prediction for the cumulative sales of Ibrance, which amounts to \$46.7 billion.

In order to gauge the precision and robustness of our proposed asset sales prediction model, we have crafted a validation designed to test the model's predictive capability under various scenarios, spanning from early stage to late stage in the product life cycle. To begin with, our selection criteria for the validation process involves picking assets that have already reached saturation in their respective markets. These assets provide an opportunity to assess our model's accuracy throughout different stages of a product's life cycle. To emulate these stages, we construct four discrete datasets based on the following key points in time: the year of saturation, the year at which 75\% of saturation was reached, the year when 50\% saturation was achieved, and the year marking 25\% saturation. These datasets effectively represent snapshots of the product’s life at different stages of its market saturation. Next, we employ a hypothetical scenario where we assume these assets have not yet reached their saturation points. This enables us to use each of the four datasets as an input to our proposed model to forecast expected sales up to the most recent available data point. Finally, acknowledging the temporal aspect of the data, we also calculate the likelihood of each asset reaching its saturation point in each subsequent year leading up to the expiration of its patent. This is done by considering the market dynamics and other pertinent factors. These calculated probabilities are then utilized to adjust the forecasted sales, with the probability for each year being multiplied by the total projected sales for that corresponding year. Aggregate estimate of expected sales are included as the sum the adjusted predicted sales from each year. This composite figure, representing our model's prediction of total sales, is then compared with the actual recorded total sales.

\begin{table}[H]
\centering
\begin{tabular}[t]{lrrrr}
\toprule
 Assets & \multicolumn{4}{c}{Differences(\%)}  \\
\midrule
\% Saturation  & 25\% &  50\% &  75\% & 100\% \\
\midrule
Prevnar family & -13.4  (19.9\%) & -11.4 (17.0\%) & 2.12 (3.15\%) & 0.371 (0.550\%)\\
Ibrance & -6.89 (22.2\%)  & -2.93 (9.45\%) & -0.114 (0.360\%)  & 0.225 (0.720\%) \\

Xeljanz       & -12.1 (86.6\%) & -8.96 (64.3\%) & -1.29 (9.23\%) & 0.376 (2.70\%) \\ 

Chantix/Champix   & -3.76 (31.4\%)  & 0.837 (6.97\%)& -11.9 (99.4\%) & 0.946  (7.88\%) \\ 

Enbrel        & 28.4 (81.8\%) & 9.26 (26.7\% )  & 9.65 (27.8\%) & 11.1 (31.8\%) \\ 

Sutent        & -6.74 (43.2\%)  & -7.33 (47.0\%)  & -10.2(65.6\%)  & 2.92 (18.7\%)  \\ 
Premarin family   & 9.05 (76.8\%)  & 1.97 (16.7\%)  & 2.09 (17.7\%)  & 2.23 (18.9\%) \\ 
Inflectra/Remsima & -1.58 (42.1\%) & -0.711 (19.0\%)   & 0.367 (9.78\%)   & 0.0510 (1.35\%)  \\ 
Xalkori       & -2.86 (56.5\%)  & -0.299 (5.90\%)   & 0.184 (3.64\%)   & 0.146 (2.88\%)  \\ \hline
{\bf Portfolio}      &\ \ -9.82 (-5.03\%)& -19.6 (-10.0\%) & -9.15 (-4.69\%)   & 18.3 (9.38\%)   \\ 

\bottomrule
\end{tabular}
\caption{\label{table:diff} Differences (in billion dollars) and \% differences for assets under four saturation levels }
\end{table}

Table \ref{table:diff} shows the difference between the estimated sales and the actual sales at saturation, after being trained on the first 25\%, 50\%, 75\%, and 100\% of data for the respective indication along with how far off they are from the actual sales expressed as a percentage of the actual sales. The last row of the table shows the same estimates aggregated over the entire portfolio. Roughly, the accuracy as a percentage of actual sales decreases in each of the indications as they are trained on more data. It can also be seen that Chanitix/Champix along with Sutent see large inaccuracies when trained on 75\% of data. For Chantix, this is likely due to a recall in 2023 \cite{lang2023}. For Sutent, it is likely because this is a relatively croweded indication space with multiple bio-similar competing for patient-share. When aggregated to the portfolio, the accuracy is at most 10\%.

\section{Post-hoc analyses} \label{sect:posthoc}

In this section we examine two aspects of the model. In the first we determine whether the sales curve could be more accurately predicted by adding auxiliary data. If this were the case, this would imply that those auxiliary data contain information that has not been incorporated into the existing sales data and that the model would be enhanced from those data. In the second, we categorize sales in the growth curve according to the disease class by evaluating the association between the category and the $\beta_1$ parameter in Equation \ref{eqn:model} for all of the therapies in the portfolio. 

\subsection{Does auxiliary asset information increase the predictive accuracy?}
\label{sect:auxiliary}

With the model's accuracy assesssed, we can ask the question ``auxiliary information increase its predictive accuracy?'' If so, this would indicate that the sales curve itself does not contain this information. While it is not possible to test the integration of all possible potential sources of predictive information, we can investigate that are likely informative: (1) the number of phase 3 trials conducted for each drug - the size of the drug program; (2) the disease type targeted by each drug. Again, we have excluded two COVID-related drugs due to their unique specialty; and (3) the number of specific indications for which the therapy is prescribed.

If any of these features were associated with the residual of the saturation value or the residual of the time to reach the saturation year, as predicted by the model, then we would conclude that the feature contains information that has not been captured by the model and those factors could be incorporated to create a more accurate model.

The described features were obtained from clinicaltrials.gov \cite{clinicaltrialsdotgov} by September $28^{th}$, 2023, focusing on interventions/treatments involving our specific drugs in the sales data. The first feature, the number of trials, were counted as trials classified as Phase 3 and falling under the category of interventional studies. The second feature took the indication target and categorized them into three distinct groups: Cancer, Infectious diseases, and others. The third, total number of conditions listed across all phase 3 trials for each drug.

To test the association between the residuals of the saturation values and the number of trials, the saturation values and the indication category, and the number of conditions targeted by the therapy, linear regressions were fit (with intercepts), a visual inspection was performed on the residuals (via Q-Q plot) to verify that they were ``roughly linear'', and the significance of the slope coefficients were examined. The results are shown for these regressions, as well as the case where the residual saturation year was the outcome, in Table \ref{tab:saturation}. None of the coefficients were significant and we fail to reject the hypothesis that these features provide predictive information beyond what is available in the estimate of the sales curve.

\begin{table}[H]
    \centering
    \begin{tabular}[t]{llll}
    \toprule
     Regresssion & Feature & Coefficients & P-value \\
    \midrule
    Residual Sales Saturation $\sim$ Number of Trials & Number of Trials & 53.8 & 0.191 \\ \hline
    Residual Sales Saturation $\sim$ Category & Cancer & -686 & 0.649 \\
     & Immune & -1600 & 0.327 \\
    Residual Sales Saturation $\sim$ Number of Cond's & Conditions & 208 & 0.134 \\ \hline
    Residual Saturation Year $\sim$ Number of Trials & Number of Trials & 0.0605 & 0.137\\ \hline
    Residual Saturation Year $\sim$ Category & Cancer & -2.00 & 0.248\\
     & Immune & -3.00 & 0.103\\ \hline
    Residual Saturation Year ~ Number of Cond's & Conditions & 0.142 & 0.340\\
    \bottomrule
    \end{tabular}
    \caption{Saturation value vs. Three Features and Saturation year vs. Three Features.}
    \label{tab:saturation}
\end{table}

\subsection{Is the indication category associated with how quickly saturation is reached?}

While the category features do not provide information that allows us to better predict future sales, one may want to examine whether category is associated with how quickly sales saturation is reached. Doing so would allow us to characterize expected sales curves by indication type.

The $\beta_1$ parameter in Equation \ref{eqn:model} is an indication specific estimate of how quickly saturation is reached with smaller values indicating a slower ``ramp-up'' to saturation and larger values indicating the converse. To test for an association between category and $\beta_1$ we perform a linear regression with $\beta_1$ as the outcome and category as the feature. The results are shown in Table \ref{tab:beta1} with "Other" as the control. The results show that immune therapies have slower ramp-up when compared to other therapy categories as indicated by a significant, negative slope coefficient. This may be because immune disease are chronic when compared to others, like cancer, and patients tend to migrate to new therapies, from ones that are perceived effective, slowly.

\begin{table}[H]
    \centering
    \begin{tabular}[t]{llll}
    \toprule
    Regression & Predictors   & Coefficients & P-value \\
    \midrule
    $\beta_1$ $\sim$ Category   & Cancer & -0.185 & 0.419  \\
         & Immune & -0.650 & 0.0172* \\
    \bottomrule
    \end{tabular}
    \caption{Regression results between $\beta_1$ and the three features}
    \label{tab:beta1}
\end{table}

\begin{figure}[htbp]
  \centering
  \begin{subfigure}[b]{0.25\textwidth}
    \includegraphics[width=\textwidth]{FinalPlots/Eliquis.pdf}
    \caption{Eliquis}
    \label{fig:subplot1}
  \end{subfigure}
  \hfill
  \begin{subfigure}[b]{0.25\textwidth}
    \includegraphics[width=\textwidth]{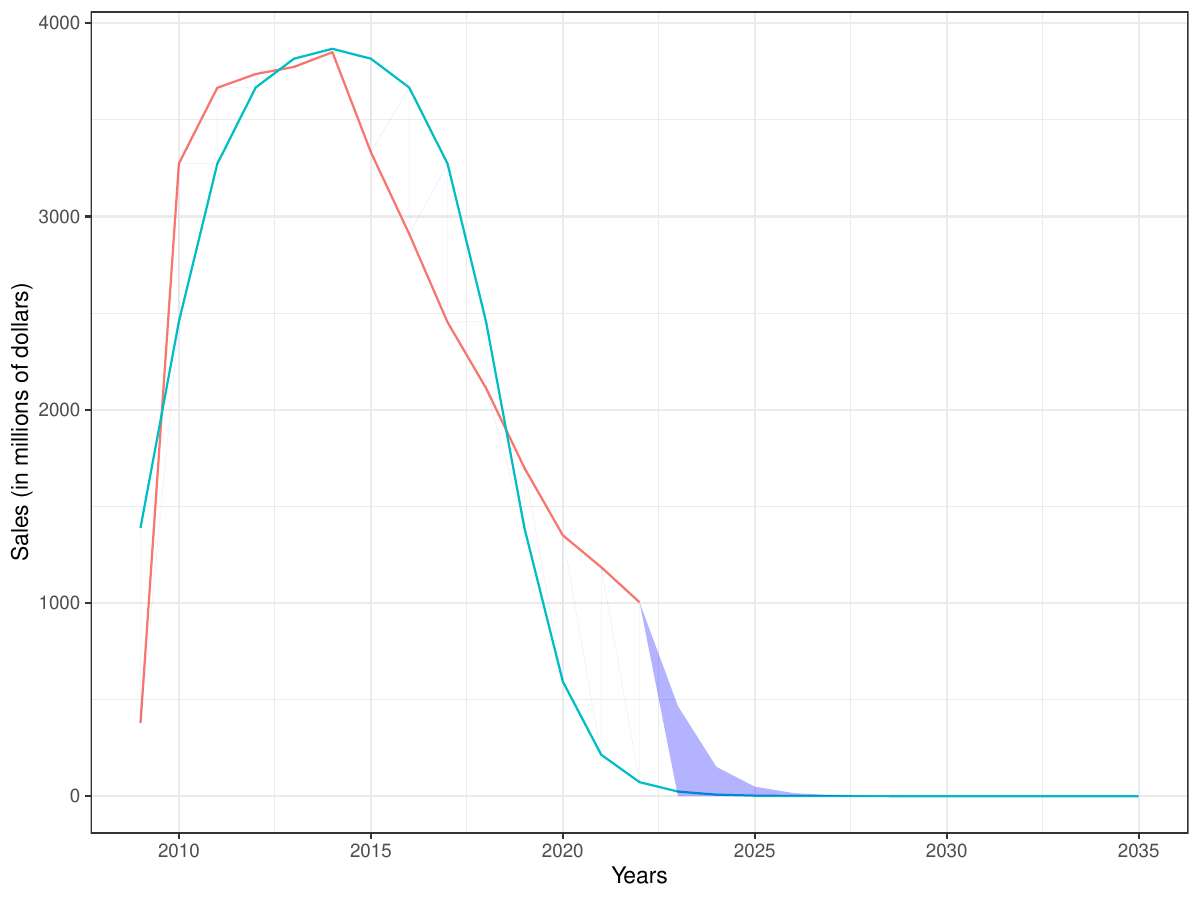}
    \caption{Enbrel}
    \label{fig:subplot2}
  \end{subfigure}
  \hfill
  \begin{subfigure}[b]{0.25\textwidth}
    \includegraphics[width=\textwidth]{FinalPlots/Ibrance.pdf}
    \caption{Ibrance}
    \label{fig:subplot3}
  \end{subfigure}
  
  \vspace{\baselineskip}
  
  \begin{subfigure}[b]{0.25\textwidth}
    \includegraphics[width=\textwidth]{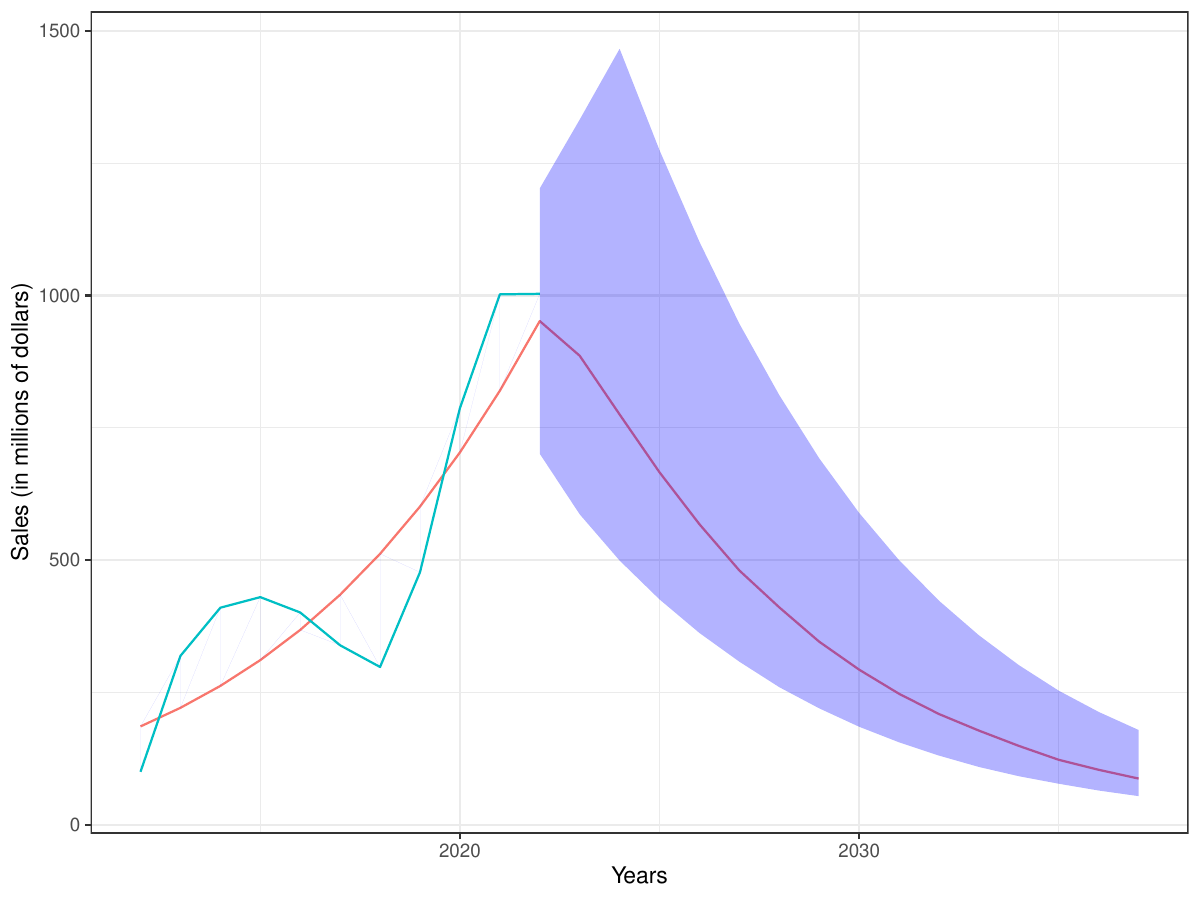}
    \caption{Inlyta}
    \label{fig:subplot4}
  \end{subfigure}
  \hfill
  \begin{subfigure}[b]{0.25\textwidth}
    \includegraphics[width=\textwidth]{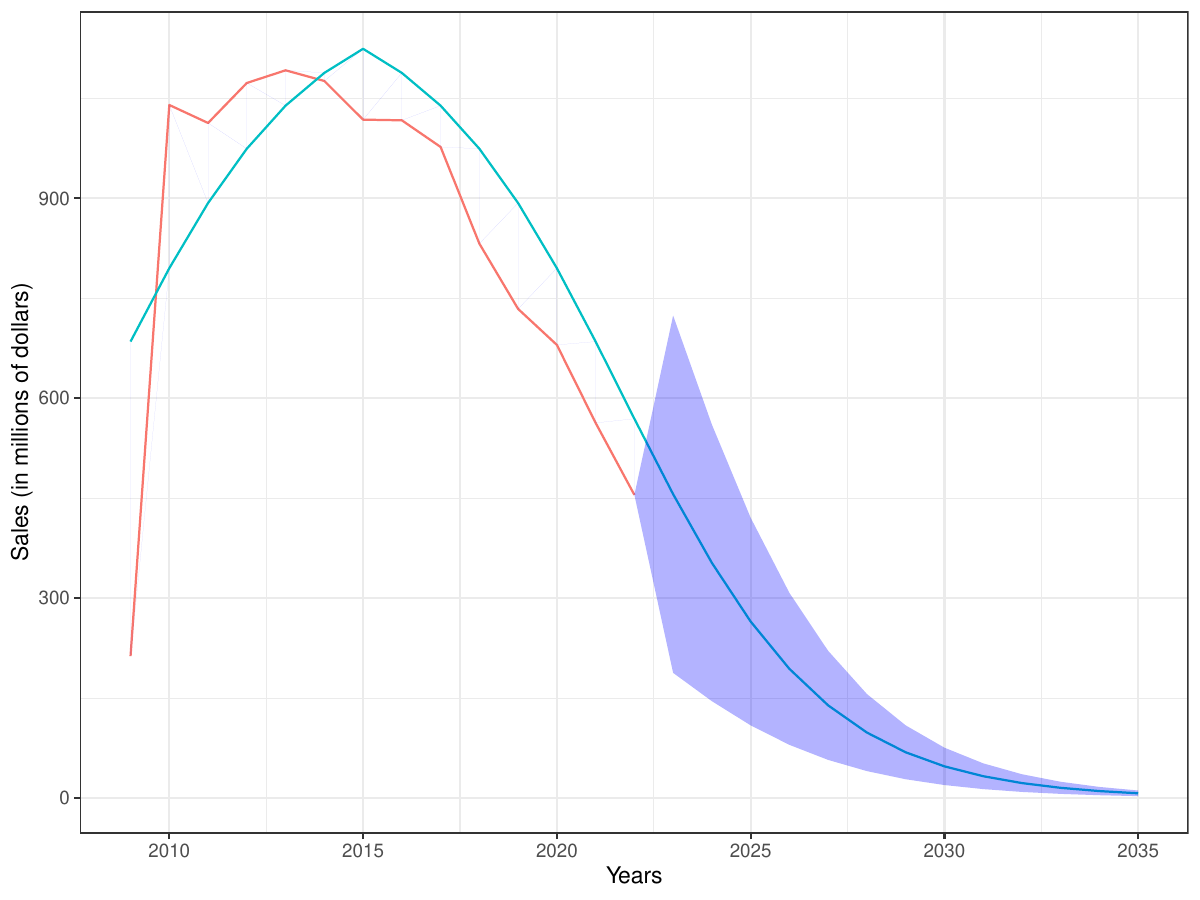}
    \caption{Premarin family}
    \label{fig:subplot5}
  \end{subfigure}
  \hfill
  \begin{subfigure}[b]{0.25\textwidth}
    \includegraphics[width=\textwidth]{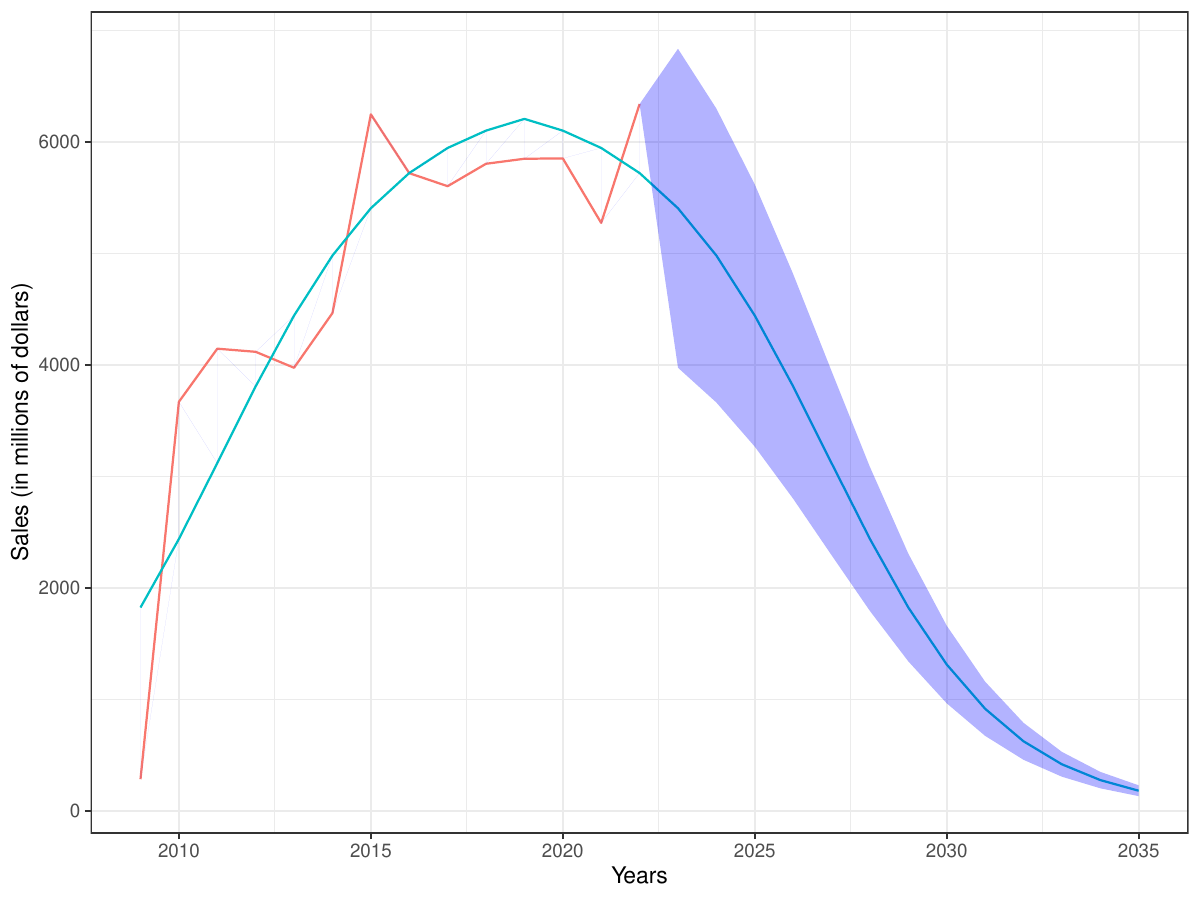}
    \caption{Prevnar family}
    \label{fig:subplot6}
  \end{subfigure}
  
  \vspace{\baselineskip}
  
  \begin{subfigure}[b]{0.25\textwidth}
    \includegraphics[width=\textwidth]{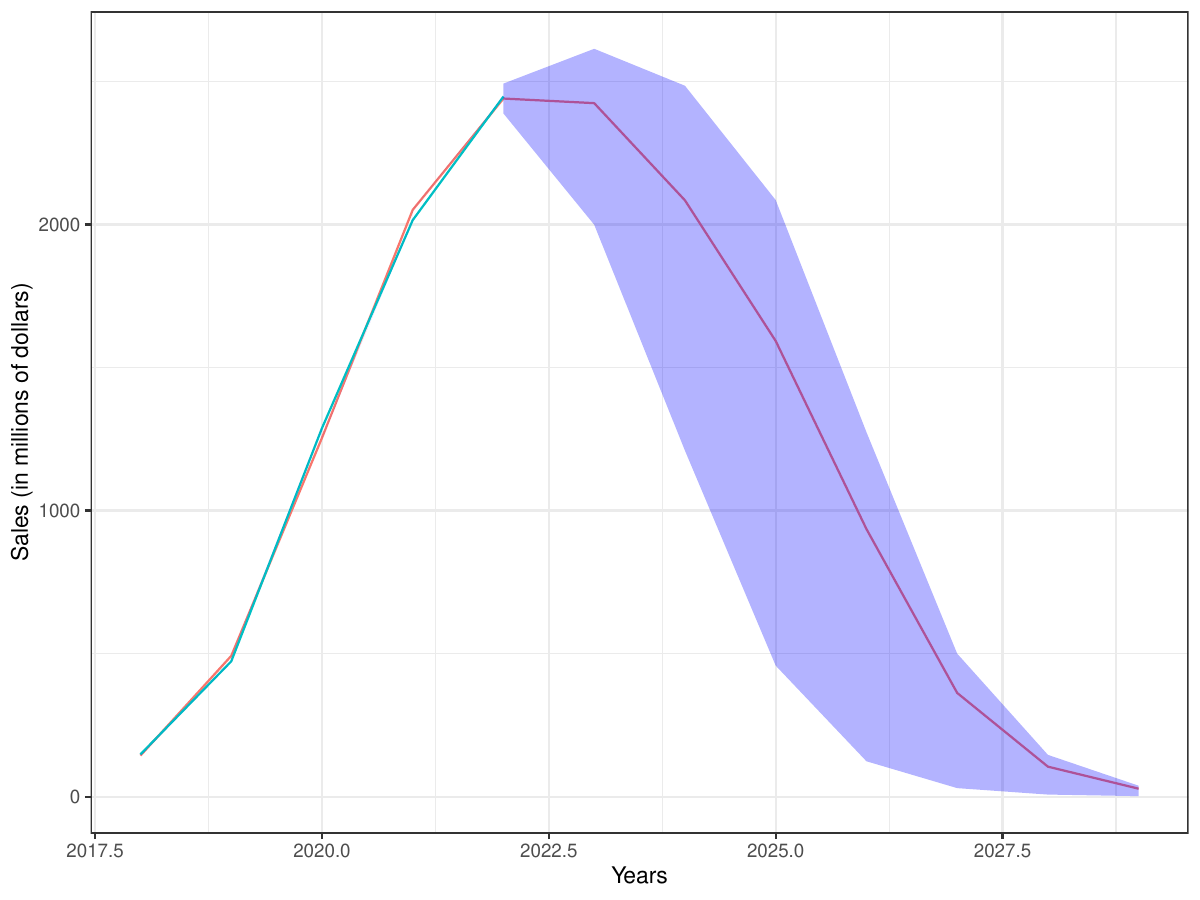}
    \caption{Vyndaqel family}
    \label{fig:subplot7}
  \end{subfigure}
  \hfill
  \begin{subfigure}[b]{0.25\textwidth}
    \includegraphics[width=\textwidth]{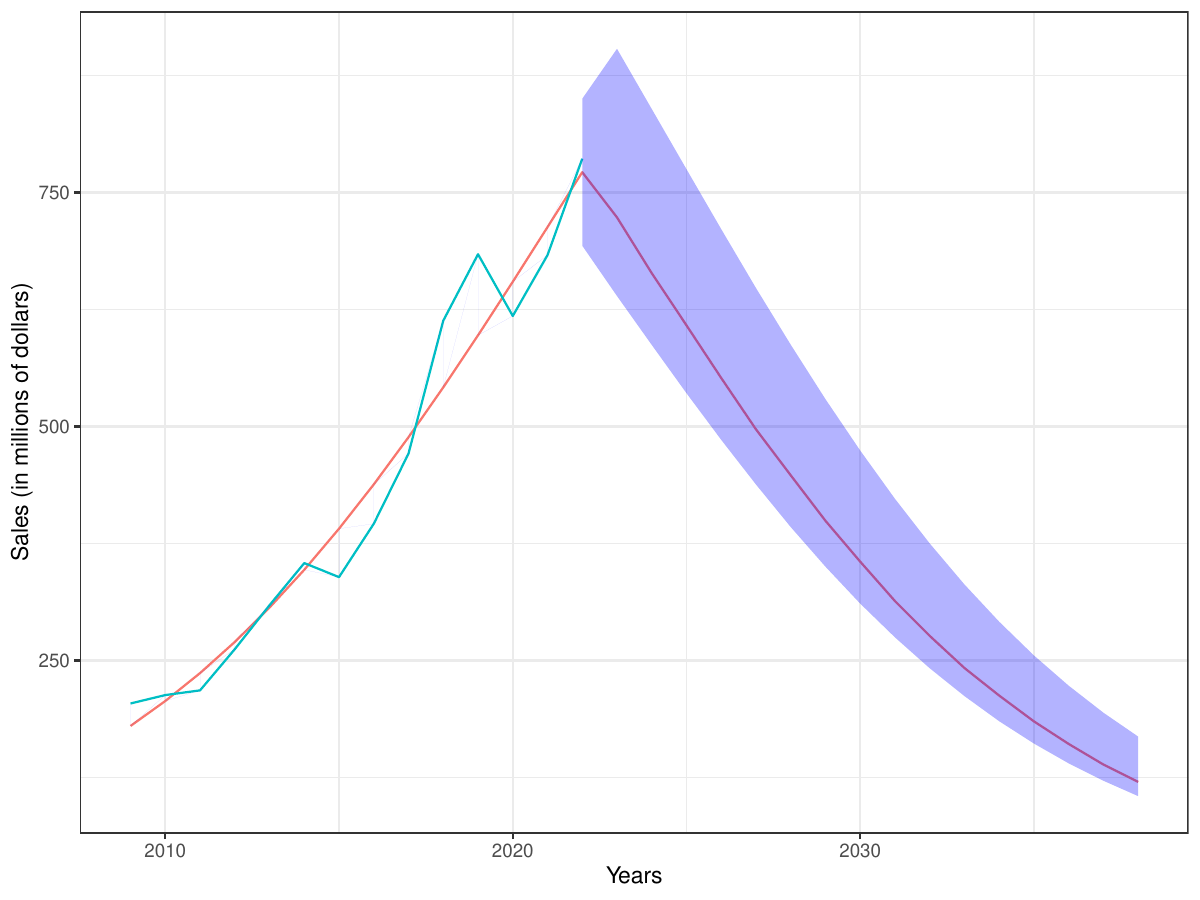}
    \caption{Sulperazon}
    \label{fig:subplot8}
  \end{subfigure}
  \hfill
  \begin{subfigure}[b]{0.25\textwidth}
    \includegraphics[width=\textwidth]{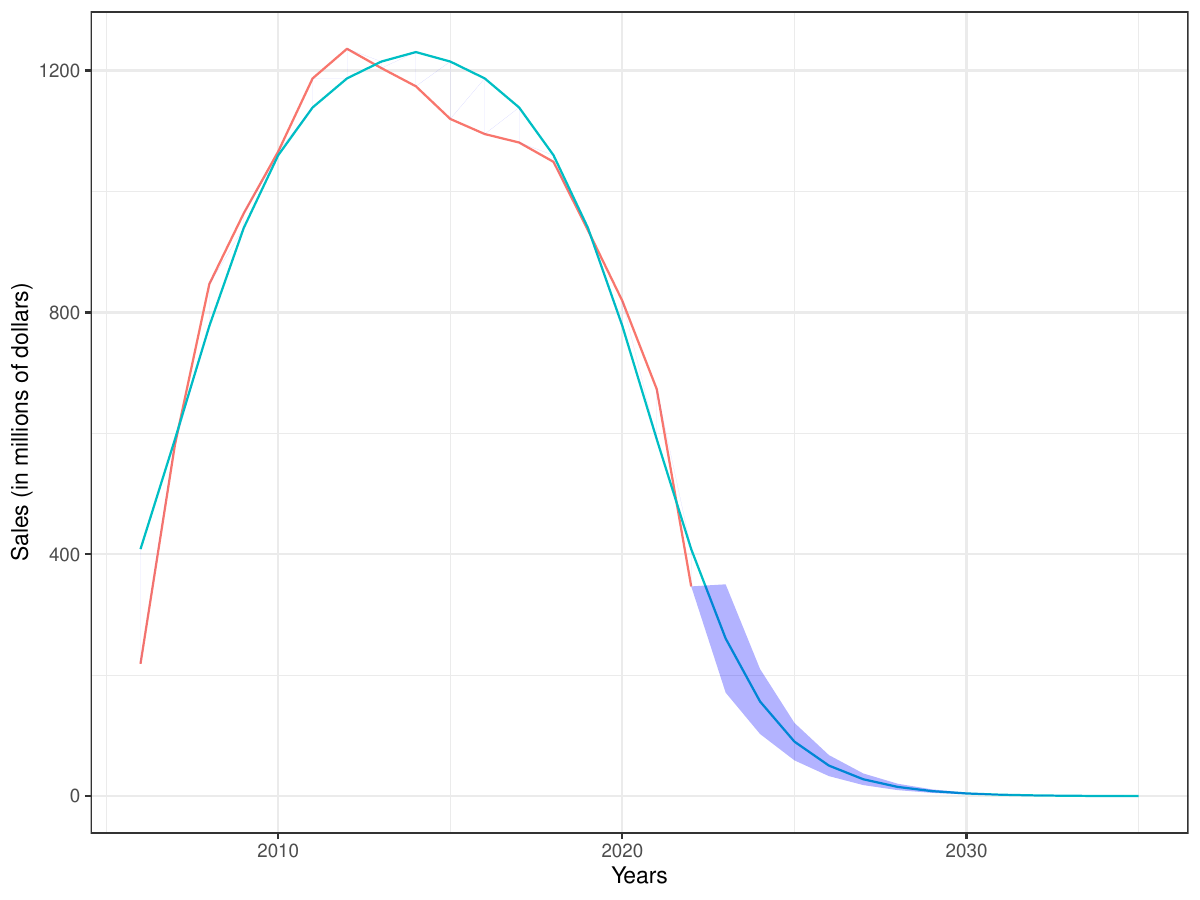}
    \caption{Sutent}
    \label{fig:subplot9}
  \end{subfigure}

  \vspace{\baselineskip}

    \begin{subfigure}[b]{0.25\textwidth}
    \includegraphics[width=\textwidth]{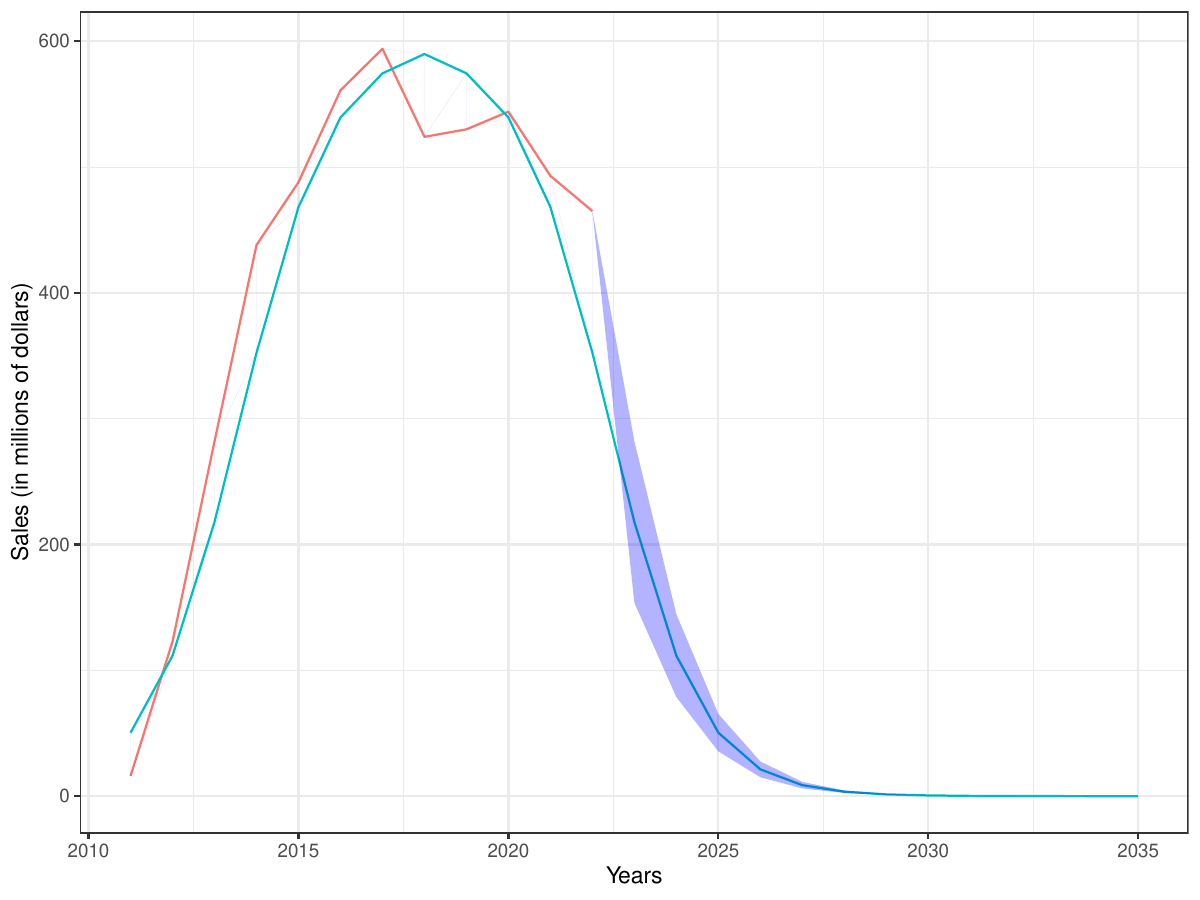}
    \caption{Xalkori}
    \label{fig:subplot10}
  \end{subfigure}
  \hfill
  \begin{subfigure}[b]{0.25\textwidth}
    \includegraphics[width=\textwidth]{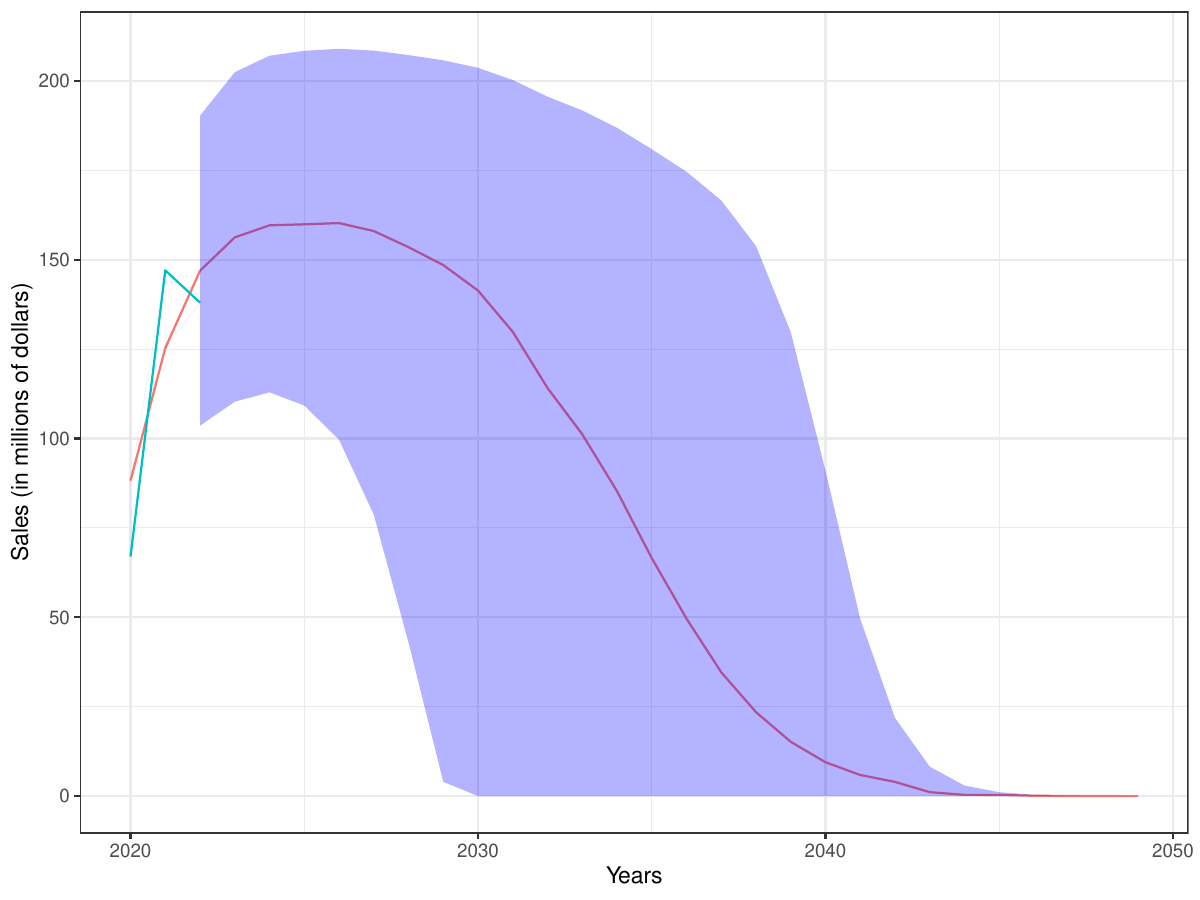}
    \caption{Eucrisa}
    \label{fig:subplot11}
  \end{subfigure}
  \hfill
  \begin{subfigure}[b]{0.25\textwidth}
    \includegraphics[width=\textwidth]{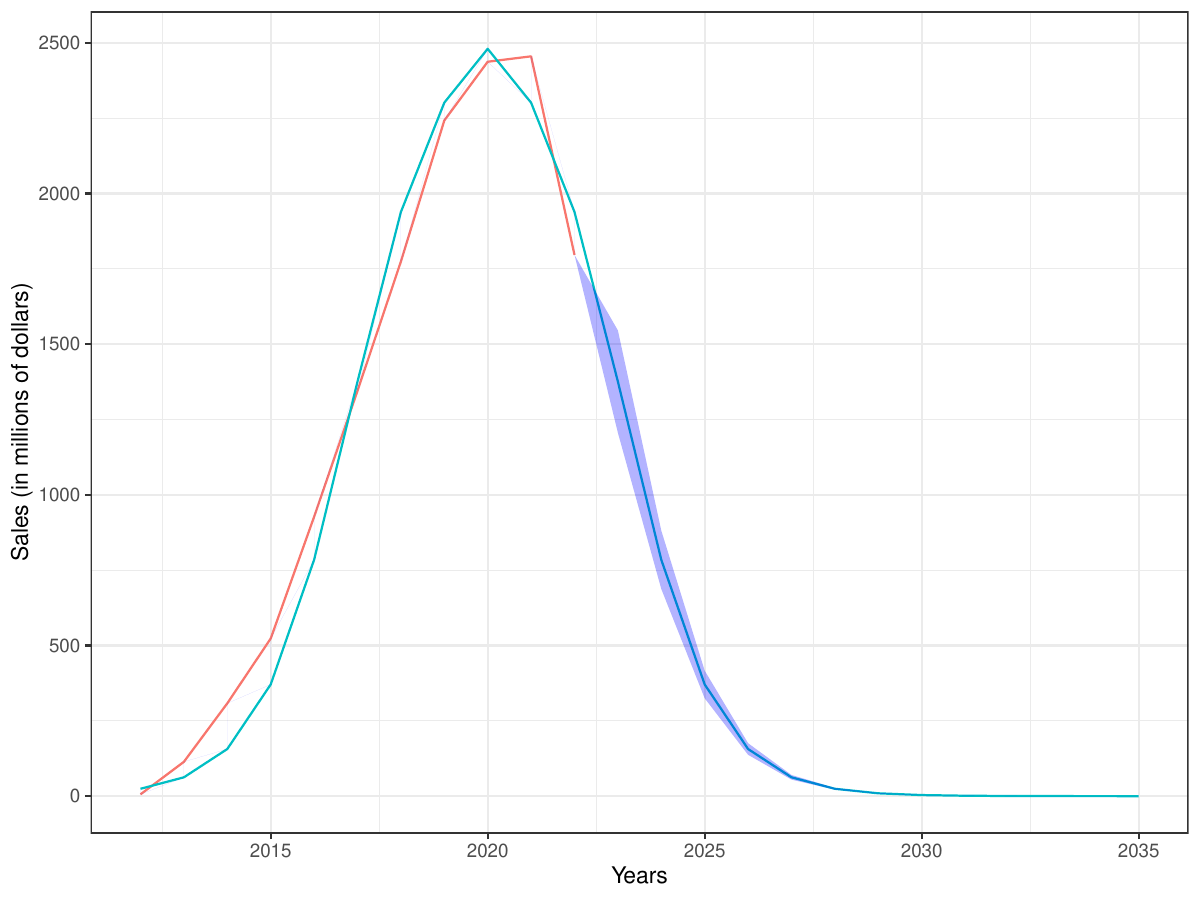}
    \caption{Xeljanz}
    \label{fig:subplot12}
  \end{subfigure}

 \vspace{\baselineskip}
 
    \begin{subfigure}[b]{0.25\textwidth}
    \includegraphics[width=\textwidth]{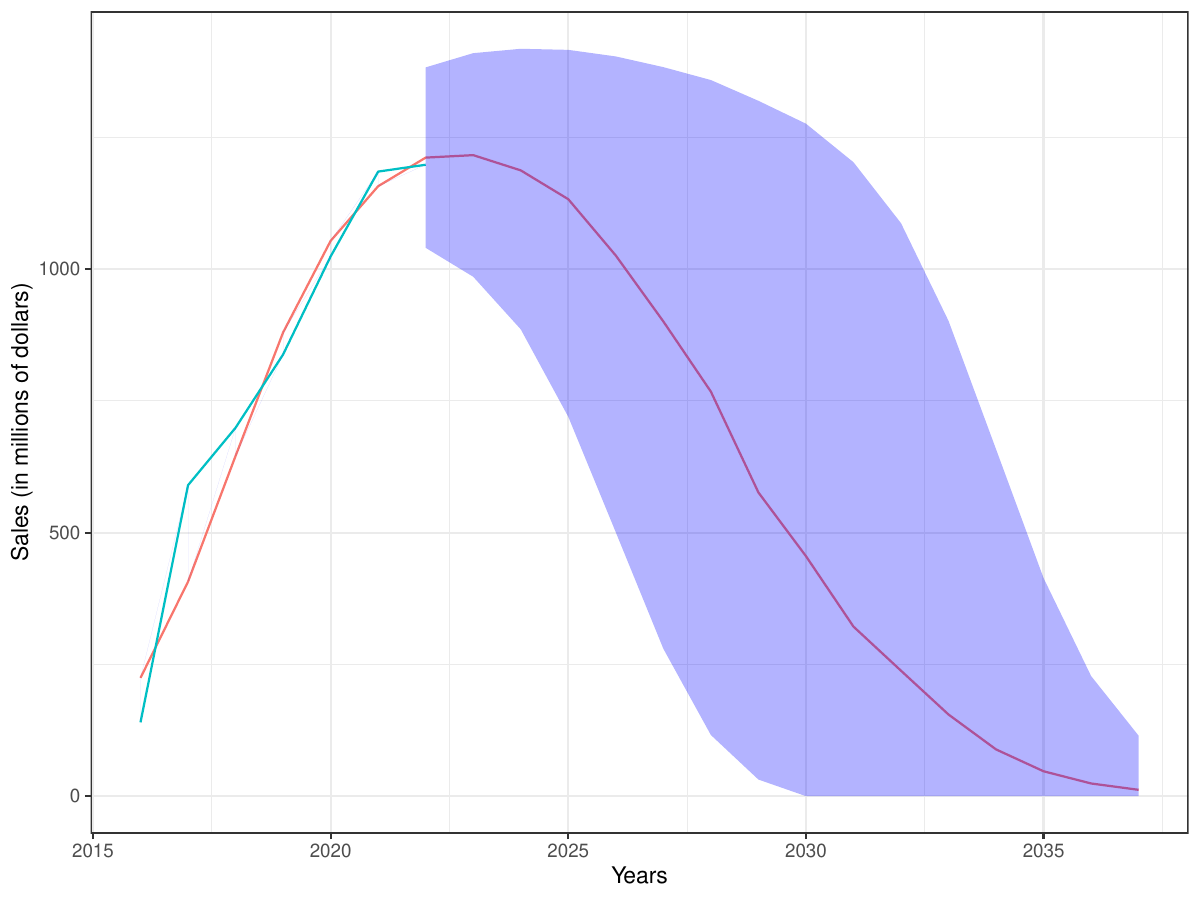}
    \caption{Xtandi}
    \label{fig:subplot13}
  \end{subfigure}
  \hfill
  \begin{subfigure}[b]{0.25\textwidth}
    \includegraphics[width=\textwidth]{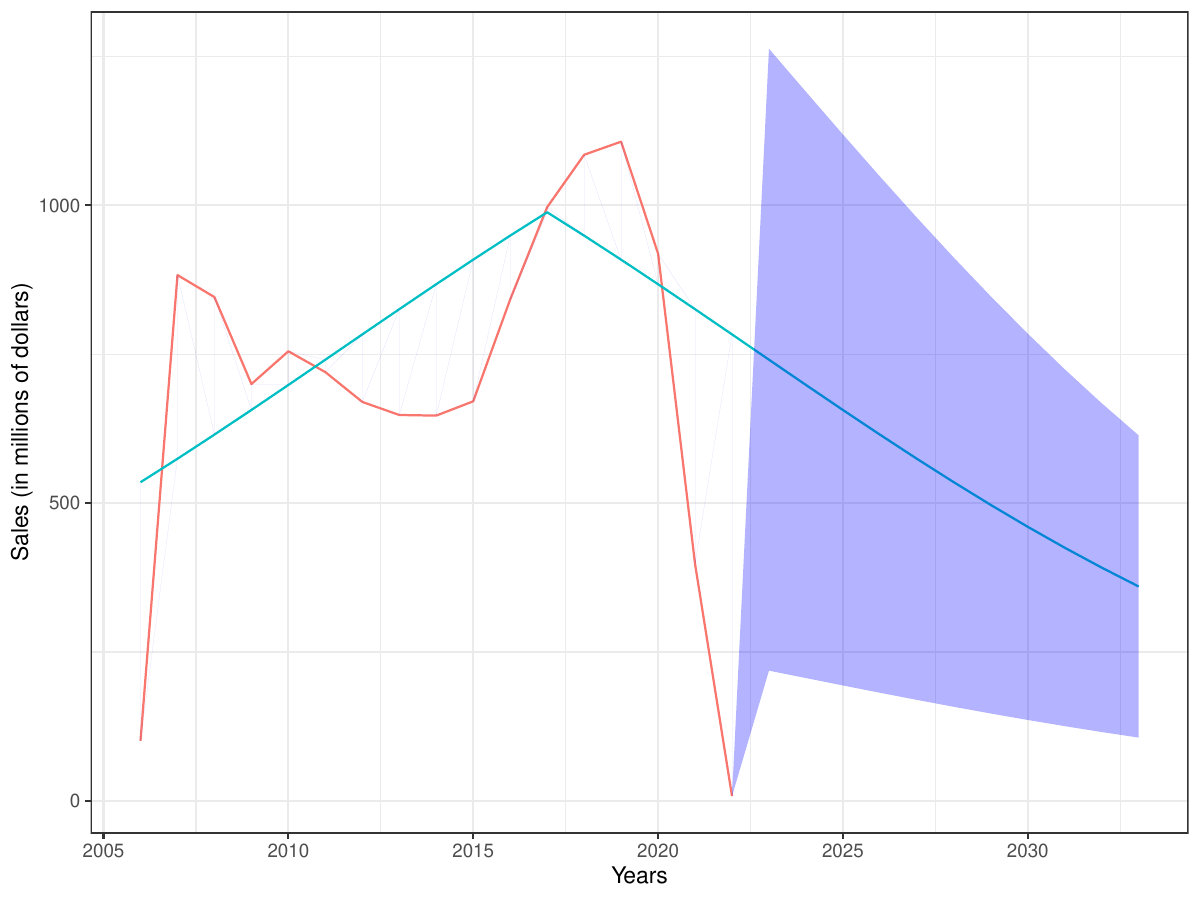}
    \caption{Champix}
    \label{fig:subplot14}
  \end{subfigure}
  \hfill
  \begin{subfigure}[b]{0.25\textwidth}
    \includegraphics[width=\textwidth]{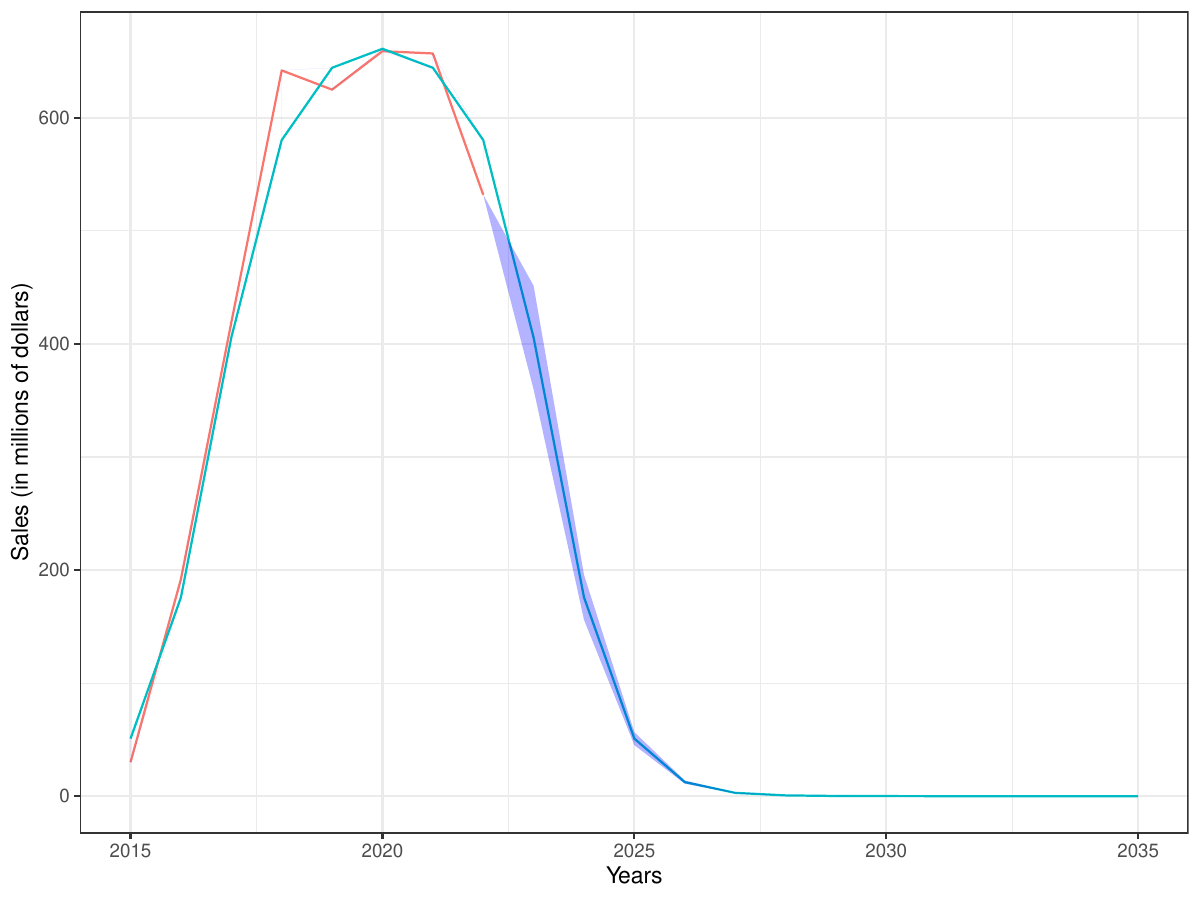}
    \caption{Remsima}
    \label{fig:subplot15}
  \end{subfigure}
  
  \caption{Figure with 15 subplots}
  \label{fig:subplots}
\end{figure}

\section{Estimating the portfolio value} \label{sect:portfolio}

Section 6 showed that the proposed model can effectively estimate the sales curves of post-revenue pharmaceutical assets. In this section, we estimate the net present value (NPV) of the post-revenue portfolio 
\begin{equation}
NPV = \sum_{i = 1}^N \sum_{t = 1}^T \frac{X_{it}}{\left(1 + r\right)^t} - I_{it}
\label{eqn:npv}
\end{equation}
where $N$ is the number of assets, $T$ is time over which the assets continue to provide cash flow, $X_{it}$ is the cash inflow for asset $i$ at time $t$, $r$ is the cost of capital, which will be fixed at 0.1, and $I_{it}$ is the cost outflow associated with asset $i$ at time $t$. For this simulation we will assume outflow is zero. In real world scenarios this will likely be a negligible fraction of the inflow.

To estimate the distribution of future cash inflows, we sample them as realizations of future values based on the sales model, incorporating the uncertainty inherent in these estimates. This methodology allows for a comprehensive understanding of the potential financial performance of the post-revenue pharmaceutical assets. We provide a detailed visualization of these assets, along with their predicted values and 95\% confidence intervals, in Figure \ref{fig:subplots}. This figure serves to illustrate the estimated performance and associated uncertainties for each asset in the portfolio.

\begin{figure}[H]
\begin{center}
    \includegraphics[width=0.8\textwidth]{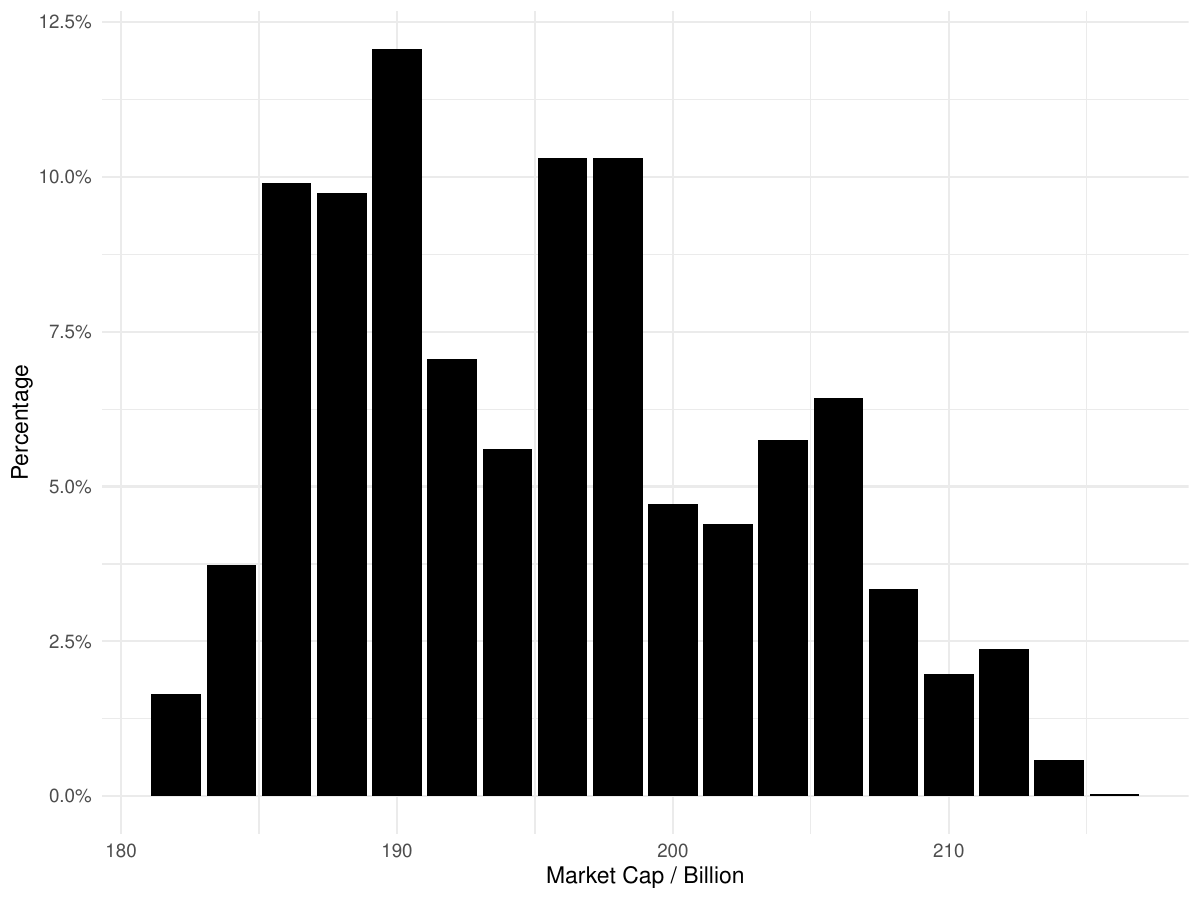}
    \caption{World-wide sales of Pfizer's Inflectra drug asset (above) and it's idealized (estimated) sales curve (below).}
\label{fig:marketcap}
\end{center}
\end{figure} 

The simulation was performed over 10000 realizations and the NPV was calculated. The distribution of sales estimates are shown in Figure \ref{fig:marketcap}. The mean price is 196.423 billion and the 95\% credible interval ranges from 185.840 to 210.383 billion. At the time this was written Pfizer's  market cap was approximately \$174 billion, its total liabilities are approximately \$117.8 billion, its current assets are worth \$74.0 billion, and it has approximately \$40 billion cash-on-hand \cite{financials}. If we take the value to be
\begin{align*}
Market\ Cap & = Current\ Asset\ Value + Pre\text{-}revenue\  Portfolio\  Value\ + \\
& Post\text{-}revenue\ Portfolio\ Value + Cash\ on\ Hand - Total\ Liabilities
\end{align*}
this would imply a pre-revenue portfolio value of approximately \$-17 billion. A speculator valuing Pfizer's pre-revenue portfolio at around \$35 billion or more would see this as a buy opportunity.

\section{Discussion} \label{sect:discussion}

Valuation of post-revenue drug assets is a critical to the success of biotechnology and pharmaceutical companies. We have proposed a model specific to pricing post-revenue biopharmaceutical assets market to provide accurate, interpretable estimates of future sales with estimates of uncertainty. While the model's application currently depends on some sales data to estimate the sales curve. We propose two extensions as avenues for future work to make it's use more general.

First, the model we have introduced, while currently reliant on some sales data, can be adapted to forecast the entire sales trajectory for a biopharmaceutical asset. This adaptation is particularly useful for revenue planning of assets that have received approval but have not yet generated sales. In such cases, it is feasible to use sales data from existing therapies within the same indication area as a baseline. These existing sales figures can then be modified to reflect the unique aspects of the new asset, such as its specific label. This label consideration is crucial as it defines the scope of the disease population from which revenue can be generated. By employing this approach, our model offers a structured methodology for estimating future sales, even in the absence of direct sales data from the new asset itself. This extension enhances the model's utility, making it a more versatile tool for strategic financial planning in the biopharmaceutical industry.

The second extension of our model addresses the challenge of valuing pre-revenue therapies, particularly those still undergoing the regulatory approval process. Similar to the previous extension, this approach involves estimating potential sales figures for these therapies. However, a unique aspect here is the consideration of the regulatory approval risk. For therapies that have not yet been approved, there is a significant risk that they may never reach the market due to regulatory hurdles. To account for this uncertainty, our model proposes integrating the probability of the drug's success into the net present value (NPV) calculation as outlined in Equation \ref{eqn:npv}. This risk-adjusted NPV approach would involve quantitatively assessing the likelihood of the drug receiving regulatory approval and reflecting this in the financial valuation. By doing so, we can provide a more accurate and realistic valuation of these assets, which is crucial for investment decisions and strategic planning in the biopharmaceutical industry. This method of incorporating success probability into the NPV framework offers a comprehensive way to evaluate the financial potential of pre-revenue therapies, balancing their projected market performance with the inherent risks associated with the drug development and approval process.

\bibliographystyle{abbrvnat}
\bibliography{refs}

\end{document}